\documentclass[
    aps,
    prx,
    letterpaper,
    nobalancelastpage,
    twocolumn,
    superscriptaddress,
    nofootinbib,
    longbibliography
]{revtex4-2}

\usepackage[colorlinks, citecolor=blue]{hyperref}
\usepackage{graphicx}
\usepackage{amsmath}
\usepackage{amssymb}
\usepackage[english]{babel}
\usepackage{color}
\usepackage[version=4]{mhchem}
\usepackage{dsfont}
\usepackage{multirow}
\usepackage{booktabs}
\usepackage{array}
\usepackage{stmaryrd}
\usepackage{physics}
\usepackage{bm}
\usepackage{comment}
\usepackage{braket}

\usepackage{textcomp}

\graphicspath{{./Figures/}}

\newcommand{\UIUC}{
    Department of Physics,
    The University of Illinois at Urbana-Champaign,
    Urbana, IL 61801, USA
}
\newcommand{\Duke}{
    Departments of Electrical and Computer Engineering, 
    Duke University,
    Durham, NC 27708, USA
}
\newcommand{\DQC}{
    Duke Quantum Center, 
    Duke University,
    Durham, NC 27708, USA
}
\newcommand{\UChicagoPME}{
    Pritzker School of Molecular Engineering,
    University of Chicago, 
    Chicago, IL 60637, USA
}
\newcommand{\UChicagoPhysics}{
    Department of Physics and James Frank Institute, 
    University of Chicago, 
    Chicago, IL 60637, USA
}
\newcommand{\Argonne}{
    Physics Division, 
    Argonne National Laboratory, 
    Lemont, IL 60439, USA
}

\begin{document}

\title{Detuning-robust Rydberg entangling gates from echoed pulses}

\author{Zhubing Jia}
\thanks{These authors contributed equally to this work. Correspondence: \href{mailto:zjia11@illinois.edu}{zjia11@illinois.edu}, \href{mailto:yichao.yu@duke.edu}{yichao.yu@duke.edu}}
\affiliation{\UIUC}

\author{Yichao Yu}
\thanks{These authors contributed equally to this work. Correspondence: \href{mailto:zjia11@illinois.edu}{zjia11@illinois.edu}, \href{mailto:yichao.yu@duke.edu}{yichao.yu@duke.edu}}
\affiliation{\Duke}
\affiliation{\DQC}

\author{Xiye Hu}
\affiliation{\UIUC}
\author{Lintao Li}
\affiliation{\UIUC}
\author{Jacob Zheng}
\affiliation{\UIUC}
\author{Christopher Monroe}
\affiliation{\Duke}
\affiliation{\DQC}
\author{Jacob P. Covey}\email{jcovey@uchicago.edu}
\affiliation{\UChicagoPME}
\affiliation{\UChicagoPhysics}
\affiliation{\UIUC}
\affiliation{\Argonne}

\begin{abstract}
Rydberg-based two-qubit gate fidelities in neutral atom arrays are limited chiefly by laser intensity inhomogeneity and by detuning errors from frequency miscalibration, background-field drift, intermediate state light shift and Doppler shifts. Quantum optimal control can suppress the former, but no pulse can render a controlled-Z gate first-order insensitive to detuning. Here we show that this obstruction can be circumvented by designing Rydberg gate pulses with every leading-order detuning error relegated to single-qubit Z-rotations, which an echoed  sequence removes. The resulting maximally-entangling ZZ gate has no leading-order response to arbitrary detunings on either atom, maintaining an infidelity below $10^{-3}$ within a much wider range of single-atom detunings than the time-optimal gates. We further show that a large fraction of residual error is dominated by population outside the computational subspace, which can be converted into heralded erasures. Our results significantly reduce the requirements on laser frequency stability and intensity homogeneity, field calibration and atomic temperature toward practical fault-tolerant quantum computing with neutral atoms.
\end{abstract}
\maketitle

\section{Introduction}\label{Intro}

Neutral atoms in optical tweezer arrays have emerged as a leading platform for quantum computation and simulation \cite{Saffman2010quantum, endres2016atom, bernien2017probing, Browaeys2020many, Henriet2020quantum, Morgado2021quantum, ebadi2021quantum, semeghini2021probing, young2022tweezer, Bluvstein2022quantum, Bluvstein2024logical, reichardt2024logical, Muniz2025Repeated, zhou2025low}. Strong, controllable interactions between Rydberg states enable two-qubit entangling gates \cite{Jaksch2000, Levine2019, Jandura2022, Ma2023, Muniz2025, senoo2026high}, and recent experiments have demonstrated programmable circuits on hundreds of atoms and logical operations on error-corrected qubits \cite{Bluvstein2022quantum, Bluvstein2024logical, Evered2023, reichardt2024logical, Muniz2025Repeated, zhang2026logical, computing2026quantum}. Entangling-gate fidelity remains among the principal limitations on scaling: while the ultimate bound is set by the Rydberg-state lifetime and the available laser power, demonstrated fidelities are dominated by technical imperfections. These are chiefly laser-intensity inhomogeneity and detuning errors arising from frequency miscalibration, background electric field drift, intermediate state light shift and Doppler shifts from finite atomic temperature \cite{de2018analysis, pagano2022error, Evered2023, Ma2023, panja2024electric, Liu2026, deng2026efficient}.

Quantum optimal control offers a route to suppressing these imperfections. Time-optimal pulses minimize the Rydberg dwell time and hence the decay error \cite{Jandura2022}, while robust pulses trade insensitivity to a specified imperfection for a longer duration \cite{Jandura2023, Fromonteil2023}. Detuning errors, however, occupy a special position: Ref. \cite{Jandura2023} proved that no pulse can render a CZ gate first-order insensitive to the atomic detunings, because the geometric phase accrued in each computational sector shifts in proportion to the time that sector spends in the Rydberg state and is thus strictly positive. Robustness against detuning has therefore been pursued by controlling the error itself rather than the pulse, for instance, by reversing the sign of the Doppler shift between two halves of the gate using a counter-propagating beam and the harmonic motion of the trapped atom \cite{Jandura2023}. 

In this work, we design a two-qubit entangling gate scheme that is robust to arbitrary detuning errors. The no-go is circumvented by noticing that the impossibility of suppressing detuning errors of the CZ gate applies to each computational sector individually, and the single-qubit component of this error can nonetheless be cancelled by applying the entangling interaction in two configurations related by exchange of the qubit states $\ket{0}$ and $\ket{1}$. We design our detuning-robust gate with an echoed sequence of global single-qubit X rotations \cite{Evered2023, evered2025probing}, in which all four sectors, including $\ket{00}$, undergo the entangling dynamics. The resulting gate has no leading-order response to detuning errors, and its residual error is dominated by Rydberg population that can be converted into erasures.

\section{Level scheme and dominant error types}
\label{Sec:2}

\begin{figure}[t!]
    \centering
    \includegraphics[width=\linewidth]{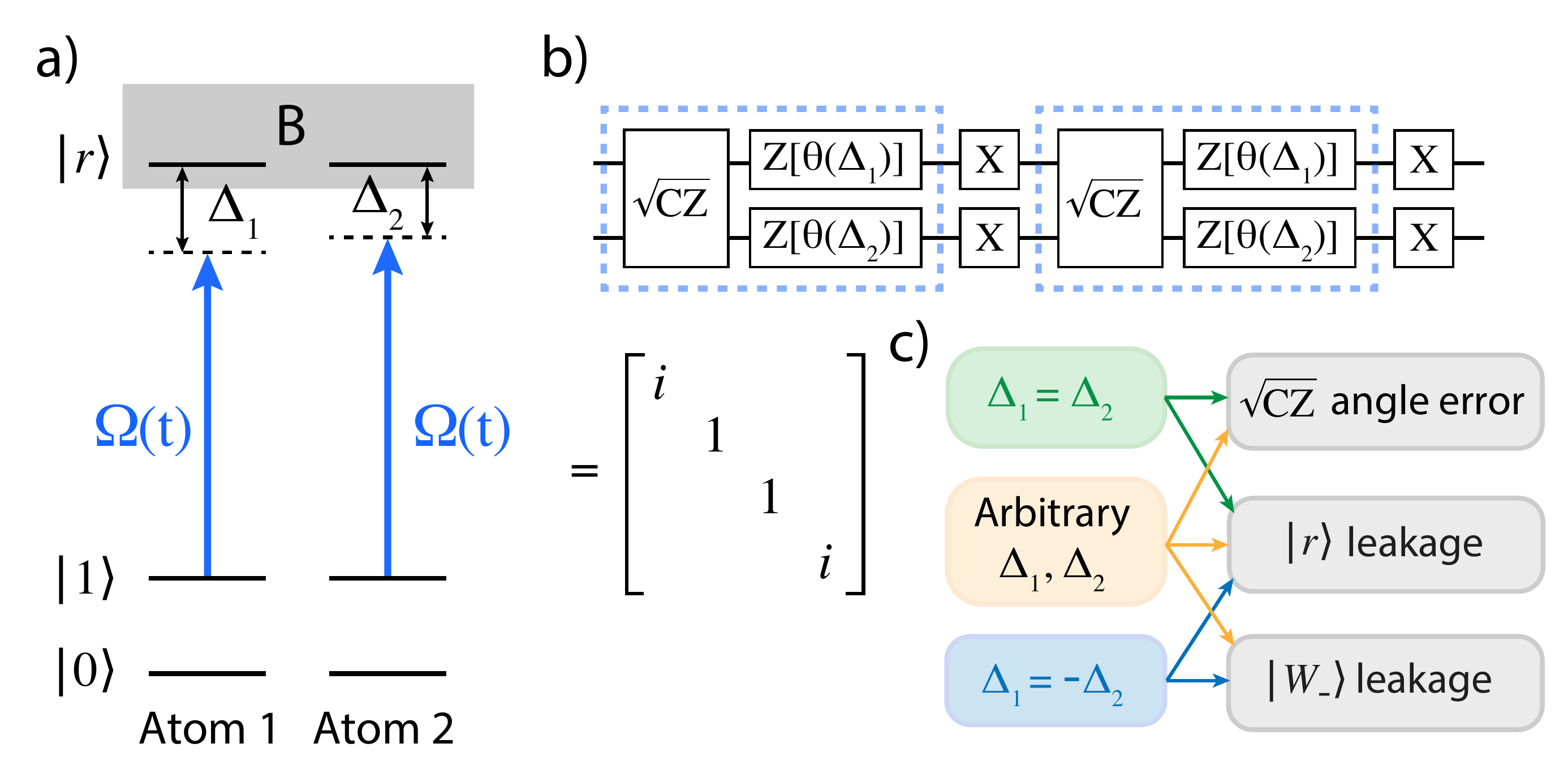}
    \caption{Overview of level structure involved in Rydberg-based entangling gates and the detuning-robust gate scheme. (a) Level scheme and transitions for the Rydberg two-qubit gate. On each atom $\ket{1}$ is coupled to $\ket{r}$ with Rabi frequency $\Omega(t)$ and unknown detuning $\Delta_{1(2)}$, with the doubly-excited state $\ket{rr}$ shifted by the van der Waals interaction $|B|\gg\Omega$. (b) Circuit for the detuning-robust entangling gate. Each dashed box denotes one Rydberg two-qubit gate comprising a $\sqrt{\text{CZ}}$ entangling operation and two single-qubit Z rotations whose angles are first-order sensitive to detuning. The interleaved single-qubit X rotations echo out these Z rotations, and the circuit implements a maximally-entangling ZZ gate. (c) Dominant error channels to be suppressed for symmetric detuning $\Delta_1=\Delta_2$, anti-symmetric detuning $\Delta_1=-\Delta_2$, and arbitrary two-atom detuning.}
    \label{figure1}
\end{figure}

We adopt the level structure standard for Rydberg-mediated entangling gates, shown in Fig.~\ref{figure1}(a). Each atom encodes a qubit in a pair of long-lived states 
$\{\ket{0}, \ket{1}\}$,
supplemented by a high-lying Rydberg state $\ket{r}$. A global, phase-modulated laser pulse $\Omega(t)=\Omega e^{i\phi(t)}$ couples $\ket{1}$ to $\ket{r}$ on both atoms. When the interatomic separation lies within the blockade radius, the van der Waals interaction shifts $\ket{rr}$ by an energy $|B|\gg \Omega$, rendering it energetically inaccessible; the two-atom dynamics is then confined to the singly-excited manifold, and the Rabi frequency of $\ket{11}\leftrightarrow\ket{W_+}$ transition, with $\ket{W_+}=\frac{\ket{1r}+\ket{r1}}{\sqrt{2}}$, is collectively enhanced by $\sqrt{2}$. The finite-blockade case is discussed in appendix \ref{sec:finiteblockade}.

In the absence of errors, the gate pulse is designed so that the $\ket{01}$, $\ket{10}$, and $\ket{11}$ states each return to themselves at the gate time $\tau$, acquiring geometric phases $\theta_{01}$, $\theta_{10}$, $\theta_{11}$ satisfying $\theta_{11}=\theta_{01}+\theta_{10}+\pi$, while $\ket{00}$ is unaffected~\cite{Levine2019, Jandura2022, Evered2023, Muniz2025}. The resulting operation is a controlled-Z (CZ) gate accompanied by two single-qubit Z-rotations. 
Using perturbation theory together with quantum optimal control, ref.~\cite{Jandura2023} established that no fully detuning-robust CZ pulse exists, essentially because the geometric phase $\theta$ cannot be made insensitive to detuning.

We instead seek pulses targeting a two-qubit rotation angle $\theta_{11}-\theta_{01}-\theta_{10}=\pi/2$ ($\sqrt{\text{CZ}}=\text{diag}[1, 1, 1, i]$ gate up to single-qubit Z-rotations) for which every leading-order detuning error is relegated to the single-qubit Z-rotations, while the two-qubit entangling phase remains robust.
The echoed sequence $\sqrt{\text{CZ}}$-X$\otimes$X-$\sqrt{\text{CZ}}$-X$\otimes$X then cancels the leading-order single-qubit Z-errors and realizes a ZZ $=\text{diag}[i, 1, 1, i]$ maximally-entangling gate that is locally equivalent to CZ up to single-qubit Z-rotations \cite{evered2025probing}, and is robust against detuning [Fig.~\ref{figure1}(b)].
We emphasize that this does not conflict with the no-go of ref. \cite{Jandura2023} as our construction leaves the geometric phase in each sector first-order sensitive, and instead makes only the combination $\theta_{11}-\theta_{01}-\theta_{10}$ stationary, while the residual single-qubit phases are removed by the echo rather than by pulse design.

We now derive the design requirements. Assuming the two atoms see detuning errors of $\Delta_1$ and $\Delta_2$ respectively, it is convenient to treat two elementary cases: symmetric detuning ($\Delta_1=\Delta_2$) and antisymmetric detuning ($\Delta_1=-\Delta_2$), since an arbitrary two-atom detuning is a linear combination of the two [Fig.~\ref{figure1}(c)]. 

For the Rydberg operations ($\sqrt{\text{CZ}}$ gates), we list the unperturbed Hamiltonian and unitary operation of the $\ket{01}, \ket{10}$ and $\ket{11}$ states:
\begin{align}
    H_{10}^{(0)}(t)&=\frac{\Omega}{2}(\ket{10}\bra{r0}e^{i\phi(t)}+\text{h.c.}), \label{eqn:H00_0}\\
    H_{01}^{(0)}(t)&=\frac{\Omega}{2}(\ket{01}\bra{0r}e^{i\phi(t)}+\text{h.c.}), \label{eqn:H01_0}\\
    H_{11}^{(0)}(t) &= \frac{\sqrt{2}\Omega}{2}(\ket{11}\bra{W_+}e^{i\phi(t)}+\text{h.c.}).\label{eqn:H11_0}
\end{align}
The unperturbed unitary operation at time $t$ has the form
\begin{align}
    U_{10/01/11}(t) = \begin{bmatrix}
    a_{10/01/11}(t) & c_{10/01/11}(t) \\
    -c^*_{10/01/11}(t) & a_{10/01/11}^*(t)
    \label{eqn:unitary}
\end{bmatrix}
\end{align}
and at gate time $\tau$,
\begin{align}
    U_{10/01/11}(\tau) = \begin{bmatrix}
    e^{i\theta_{10/01/11}} &   \\
      & e^{-i\theta_{10/01/11}}
      \label{eqn:gate}
\end{bmatrix}.
\end{align}
Because $H^{(0)}$ is invariant under exchange of the two atoms, $\theta_{01}=\theta_{10}$.

Under symmetric detuning $\Delta_1=\Delta_2=\Delta_s$, the perturbed Hamiltonian is also invariant under atom swap. The perturbed $H_{10}$, $H_{01}$ and $H_{11}$ reads
\begin{align}
    H_{01}(t) &= H_{01}^{(0)}(t) + \Delta_s\ket{0r}\bra{0r},
    \label{eq:pert_sym_01}
    \\
    H_{10}(t) &= H_{10}^{(0)}(t) + \Delta_s\ket{r0}\bra{r0},
    \label{eq:pert_sym_10}
    \\
    H_{11}(t) &= H_{11}^{(0)}(t) + \Delta_s\ket{W_+}\bra{W_+}.
    \label{eq:pert_sym_11}
\end{align}
Under anti-symmetric detuning $\Delta_1=-\Delta_2=\Delta_a$, the perturbed Hamiltonian reads
\begin{align}
    H_{01}(t) &= H_{01}^{(0)}(t) + \Delta_a\ket{0r}\bra{0r},\\
    H_{10}(t) &= H_{10}^{(0)}(t) - \Delta_a\ket{r0}\bra{r0},\\
    H_{11}(t) &= H_{11}^{(0)}(t) + \Delta_a(\ket{W_+}\bra{W_-}+\text{h.c.})
    ~\label{eqn:pert_asym}
\end{align}
with $\ket{W_-}=\frac{\ket{1r}-\ket{r1}}{\sqrt{2}}$.

We now discuss the pulse design requirements for all three (symmetric, anti-symmetric, arbitrary two-atom) detuning cases.

\subsection{Controlling residual Rydberg population under detuning error}

We first consider the condition to eliminate the residual population in the Rydberg state after each Rydberg operation. Since this is a constraint on the two level evolution, we will express it using a generic two-level system $\{\ket{g'},\ket{r'}\}$ which maps to $\{\ket{01},\ket{0r}\}$, $\{\ket{10},\ket{r0}\}$ or $\{\ket{11},\ket{W_+}\}$ for each sector.
The ideal evolution $U^{(0)}(t)$ is generated by $H^{(0)}(t)=\frac{\Omega}{2}(\ket{g'}\bra{r'}e^{i\phi(t)}+\text{h.c.})$, designed such that $U^{(0)}(\tau) = \text{diag}[e^{i\theta}, e^{-i\theta}]$. Adding the detuning error $\Delta H^{(1)} = \Delta\ket{r'}\bra{r'}$ and writing $U(t) = U^{(0)}(t)+\Delta U^{(1)}(t) + O(\Delta^2)$, first-order perturbation theory gives
\begin{align}
    \bra{r'}U^{(1)}(\tau)\ket{g'} = -ie^{-i\theta}\int_0^{\tau} a(t)c(t)\text{d}t
    \label{leakage}
\end{align}
and
\begin{align}
    \bra{g'}U^{(1)}(\tau)\ket{g'} = -i e^{i\theta} \int_0^{\tau} |c(t)|^2 \text{d}t
    \label{angle}
\end{align}
with $a(t)=\bra{g'}U^{(0)}(t)\ket{g'}$ and $c(t)=\bra{r'}U^{(0)}(t)\ket{g'}$ [cf. Ref. \cite{Jandura2023}].
Equation \eqref{leakage}, the $\ket{r}$-leakage amplitude, quantifies the population that fails to return to $\ket{g'}$. We can null this error by designing the pulse so that
\begin{align}
    \bra{r'}U^{(1)}(\tau)\ket{g'} \propto \int_0^{\tau} a(t)c(t)\text{d}t=0
    \label{leakagecondition}
\end{align}
for all sectors.

With the $\ket{r}$-leakage nulled, the residual first-order effect \eqref{angle} is a purely geometric phase error, changing ${\theta}$ to ${\theta - \Delta \int_0^{\tau} |c(t)|^2 \text{d}t}$.

\subsection{Robust pulse design for symmetric detuning error}
\label{SDR}

Under $\Delta_1=\Delta_2=\Delta_s$ each sector of $\left\{\ket{01},\ket{10},\ket{11}\right\}$ acquires a detuning on its Rydberg component [Eqs. \eqref{eq:pert_sym_01}\eqref{eq:pert_sym_10}\eqref{eq:pert_sym_11}], reducing to a detuned two-level problem. Once the $\ket{r}$-leakage in all three sectors 
are suppressed, symmetric-detuning robustness of the entangling phase requires the first-order derivative of $\theta_{11}-\theta_{10}-\theta_{01}=\theta_{11}-2\theta_{10}$ to vanish:
\begin{align}
\frac{\partial(\theta_{11}-2\theta_{10})}{\partial \Delta_s} = 2\int_0^{\tau}|c_{10}(t)|^2\text{d}t - \int_0^{\tau}|c_{11}(t)|^2\text{d}t = 0.
   \label{robust}
\end{align}
A pulse satisfying \eqref{leakagecondition} and \eqref{robust} confines all detuning errors to the single-qubit Z-angles while preserving a robust entangling phase under small symmetric detuning.

\subsection{Robust pulse design for anti-symmetric detuning error}
\label{ADR}

For $\Delta_1=-\Delta_2=\Delta_a$, the $\ket{01}$ and $\ket{10}$ sectors evolve under opposite detunings, so with $\ket{r}$-leakage numerically suppressed their phase errors satisfy ${\partial \theta_{01}}/{\partial\Delta_a}=-{\partial \theta_{10}}/{\partial\Delta_a}$. The evolution of $\ket{11}$ sector now couples $\ket{W_+}$ to $\ket{W_-}$ [Eq. \ref{eqn:pert_asym}]. Because this coupling is purely off-diagonal and $\ket{W_-}$ is unpopulated at zeroth order, the first-order correction has no projection onto the $\ket{11}$-$\ket{W_+}$ manifold,
\begin{align}
    \bra{11}U^{(1)}_{11}(\tau)\ket{11} = \bra{W_+}U^{(1)}_{11} (\tau)\ket{11} = 0,
\end{align}
and instead populates the $\ket{W_-}$ state,
\begin{align}
    \bra{W_-}U^{(1)}_{11}(\tau)\ket{11} = -i\int_0^\tau c_{11}(t)\text{d}t.
    \label{W-}
\end{align}
Suppressing this $\ket{W_-}$-leakage by setting
\begin{align}
    \int_0^\tau c_{11}(t)\text{d}t = 0
    \label{W-condition}
\end{align}
renders the $\{\ket{11},\ket{W_+}\}$ evolution insensitive to $\Delta_a$ at first order, ${\partial \theta_{11}}/{\partial\Delta_a}=0$. Combined with $\ket{r}$-leakage suppression and the exchange relation above, the entangling phase is then automatically robust, ${\partial(\theta_{11}-\theta_{10}-\theta_{01})}/{\partial \Delta_a}=0$. Hence a pulse satisfying \eqref{leakagecondition} and \eqref{W-condition} confines all leading-order antisymmetric-detuning errors to the single-qubit angles.

\subsection{Robust pulse design for arbitrary two-atom detuning error}

With arbitrary detuning errors on two atoms $\Delta_1$ and $\Delta_2$, the perturbed Hamiltonian reads
\begin{align}
    H_{10/01}(t) =& H_{10/01}^{(0)}(t) + \Delta_{1/2} \ket{r0/0r}\bra{r0/0r}, \\
    H_{11}(t) =& H_{11}^{(0)}(t) + \Delta_a(\ket{W_+}\bra{W_-}+\text{h.c.})\nonumber \\
    & + \Delta_s(\ket{W_+}\bra{W_+}
    +\ket{W_-}\bra{W_-})
\end{align}
with $\Delta_s=(\Delta_1+\Delta_2)/2$ and $\Delta_a=(\Delta_1-\Delta_2)/2$.
Every correction term has been treated above except $\Delta_s\ket{W_-}\bra{W_-}$, an energy shift on the $\ket{W_-}$ state. However, since this term only acts on the $\ket{W_-}$ state which has an amplitude of $O(\Delta_a)$ [Eq. \eqref{W-}], its contribution occurs at higher order of $O(\Delta_a\Delta_s)$, leaving the leading-order robustness unaffected. 

Consequently, suppressing
$\ket{r}$-leakage \eqref{leakagecondition}, two-qubit rotation angle derivative  \eqref{robust} and $\ket{W_-}$-leakage \eqref{W-condition} eliminates the leading-order corrections from all perturbation terms. The resulting $\sqrt{\text{CZ}}$ pulse carries a two-qubit entangling phase robust against arbitrary small two-atom detuning, with all leading-order errors residing on single-qubit Z-rotations, and is suitable for the echoed $\sqrt{\text{CZ}}$-X$\otimes$X-$\sqrt{\text{CZ}}$-X$\otimes$X sequence.

\section{Numerical optimization and simulation results}
\label{sec:3}

Following the fidelity measure commonly used for CZ gates on the Rydberg platform \cite{Levine2019, Theis2016, Graham2019, Robicheaux2021, Jandura2022, Jandura2023}, we quote the $\sqrt{\text{CZ}}$ fidelity as
\begin{align}
    F = \frac{1}{16}\left|1+\sum_{q\in\left\{01, 10, 11\right\}}e^{-i\theta_q}\bra{q}U(\tau)\ket{q}\right|^2
\end{align}
with $\theta_q$s satisfying $\theta_{11}-\theta_{10}-\theta_{01}=\pi/2$.

To search for $\sqrt{\text{CZ}}$ pulses under different robustness conditions, we use quantum optimal control \cite{Glaser2015} to simultaneously minimize infidelity and all perturbation terms using the cost function
\begin{align}
    J =& 1 - F \nonumber\\
    &+ \alpha\left(2\left|\int_0^{\tau} a_{10}(t)c_{10}(t)\text{d}t\right|^2 + \left|\int_0^{\tau} a_{11}(t)c_{11}(t)\text{d}t\right|^2\right)\nonumber\\ 
    &+ \beta\left(2\int_0^{\tau}|c_{10}(t)|^2\text{d}t - \int_0^{\tau}|c_{11}(t)|^2\text{d}t\right)^2\nonumber \\ 
    &+ \gamma\left|\int_0^\tau c_{11}(t)\text{d}t\right|^2
    \label{eq:costfunc}
\end{align}
where $\alpha$, $\beta$, $\gamma \geq 0$ are tunable weights 
for the $\ket{r}$-leakage, entangling-angle, and $\ket{W_-}$-leakage conditions of Eqs. \eqref{leakagecondition}, \eqref{robust}, and \eqref{W-condition}, respectively.
Following refs. \cite{Jandura2022, Jandura2023}, 
we divide $\tau$ uniformly into $N\gg 1$ segments and take $\phi(t)$ piecewise constant, $\phi(t)=\phi_j$ for $t\in [(j-1)\tau/N,j\tau/N]$. 
We use the numerical GRAPE algorithm to efficiently calculate the derivative of $J$ with respect to each $\phi_j$, and use a gradient-descent optimizer to optimize for a phase profile that minimizes $J$ \cite{khaneja2005optimal}. 

In simulating gate performance we neglect the finite lifetime of $\ket{r}$, assume ideal single-qubit X rotations, and treat the detuning errors as quasi-static over the full echoed sequence. To compare with the normal CZ gate scheme, we search for the $\sqrt{\text{CZ}}$ solution with the smallest $\Omega\tau$ while holding error-free gate fidelity $\gtrsim 99.99\%$ and benchmark against a time-optimal (TO) pulse as in ref. \cite{Jandura2022}.
We note that the echoed sequence costs twice the duration of the corresponding $\sqrt{\text{CZ}}$ gate plus the time required for the two single-qubit X rotations, which takes $\mu$s-level time using stimulated Raman transitions \cite{Levine2019, lis2023midcircuit, li2025parallelized}. Additional gate error from finite Rydberg state lifetime is discussed in appendix \ref{sec:finitelifetime}.

\subsection{Anti-symmetric detuning-robust (ADR) gate}

We first seek pulses robust against anti-symmetric detuning errors ($\Delta_1=-\Delta_2=\Delta$) by setting $\beta=0$ in Eq. \eqref{eq:costfunc}, retaining the $\ket{r}$-leakage and $\ket{W_-}$-leakage conditions required by Sec. \ref{ADR}.
We note that neither systematic nor stochastic imperfections typically produce purely anti-symmetric detuning, but the broad, flat response of the ADR gate makes it well suited to configurations with a built-in differential detuning between the two atoms, such as a velocity-induced differential Doppler shift \cite{Lib2026, Xue2026DShor} or a tweezer-induced differential Stark shift.

\begin{figure}
    \centering
    \includegraphics[width=\linewidth]{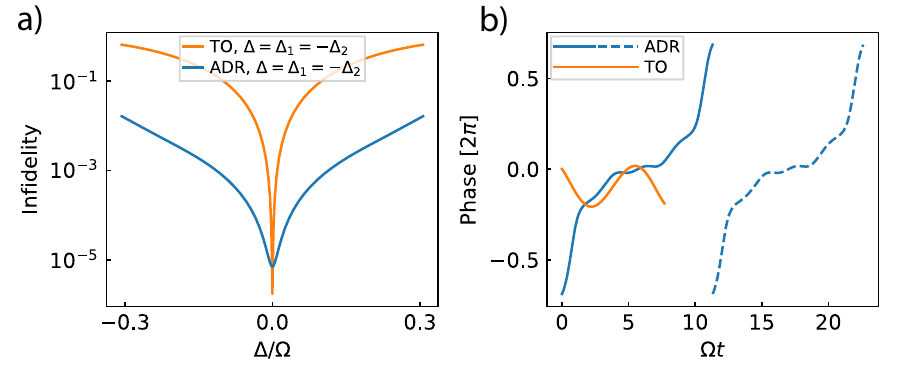}
    \caption{
    (a) Simulated gate infidelity under anti-symmetric two-atom detuning ($\Delta_1=-\Delta_2=\Delta$) for the ADR and TO pulses. (b) Optimized phase profiles of the ADR and TO pulses, with $\Omega\tau_{\text{TO}}=7.695$ and $\Omega\tau_{\sqrt{\text{CZ}}}=11.31$. The blue solid and dashed segments show the two $\sqrt{\text{CZ}}$ halves of the echoed sequence. The phase profile exhibits an approximately linear trend, corresponding to a constant detuning offset of the drive, and a similar trend appears in the phase profiles shown throughout.
    }
    \label{fig:ADR}
\end{figure}

Figure \ref{fig:ADR} shows the resulting gate infidelity under anti-symmetric detuning together with the ADR and TO phase profiles. The ADR pulse reaches an error-free infidelity at the $10^{-5}$ level and remains below $10^{-3}$ over a wide range of anti-symmetric detuning, $|\Delta|/\Omega\leq0.1$, while for TO a below $10^{-3}$ infidelity requires $|\Delta|/\Omega\leq0.01$.
Over the simulated detuning range $|\Delta|/\Omega\leq 0.3$, ADR gate outperforms the TO gate by more than one order of magnitude.

\subsection{Detuning-robust (DR) gate}

\begin{figure}[t!]
    \centering
    \includegraphics[width=\linewidth]{
    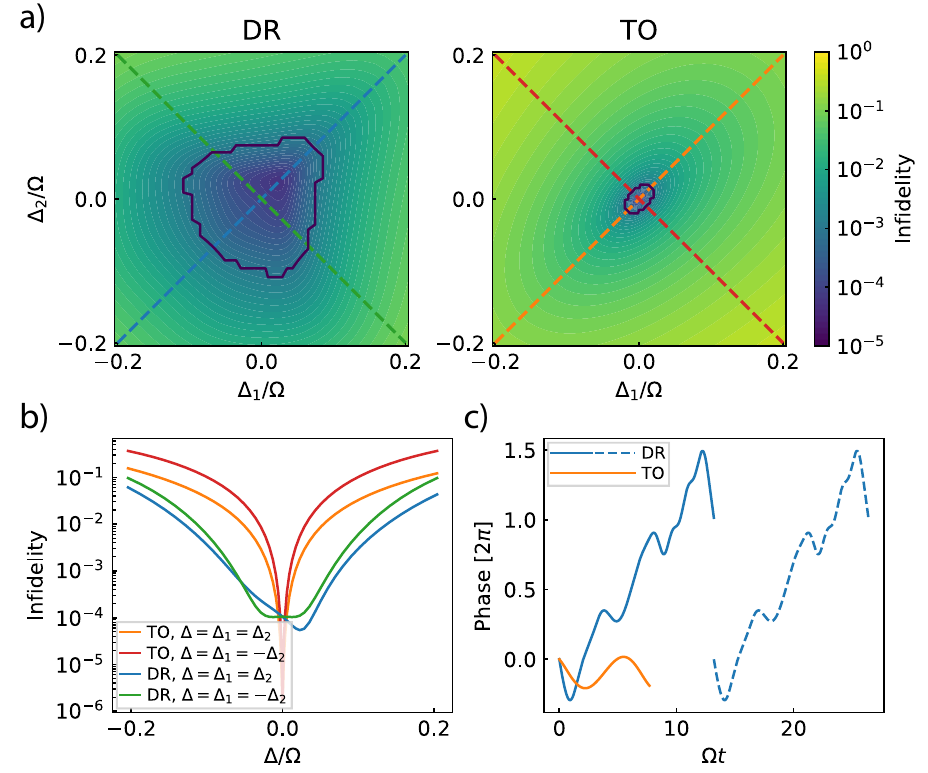}
    \caption{
    (a) Simulated gate infidelity as a function of the two-atom detunings $\Delta_1$ and $\Delta_2$ for the DR (left) and TO (right) pulses. Contours enclose the region of simulated fidelity above $99.9\%$. 
    (b) Infidelity along the symmetric ($\Delta_1=\Delta_2$) and antisymmetric ($\Delta_1=-\Delta_2$) diagonals of (a); the leading-order response to detuning is removed for the DR pulse along both directions. (c) Optimized phase profiles of the DR and TO pulses with $\Omega\tau_{\text{TO}}=7.695$ and $\Omega\tau_{\sqrt{\text{CZ}}}=13.195$. The blue solid and dashed segments show the two $\sqrt{\text{CZ}}$ halves of the echoed sequence.
    }
    \label{fig:2d-detuning}
\end{figure}

We then minimize all three terms in Eq. \eqref{eq:costfunc} to obtain pulses robust against arbitrary two-atom detuning, including but not limited to laser frequency drift (symmetric), background E-field drift (symmetric), intermediate-state Stark shift (mostly symmetric but with local variation) and thermal Doppler shift (symmetric + anti-symmetric), which we refer to as detuning-robust (DR) pulses. Figure \ref{fig:2d-detuning}(c) shows the optimized DR and TO phase profiles. Relative to the ADR pulse, the DR pulse requires a 16\% longer gate duration at fixed Rabi frequency and its phase traverses a wider range, reflecting the additional constraint. Figures \ref{fig:2d-detuning}(a) and \ref{fig:2d-detuning}(b) show the corresponding gate infidelity. The error-free infidelity of the DR pulse only reaches the $10^{-4}$ level, but can be otherwise reduced by increasing the gate time or the Rabi frequency, and the floor reflects the numerical optimization rather than a fundamental limit.
The DR pulse nonetheless eliminates the leading, quadratic dependence of the infidelity on arbitrary detunings, and maintains an infidelity below $10^{-3}$ out to single-atom detunings of $|\Delta_{i}|/\Omega\approx0.06$ ($i=1,2$), while for TO an infidelity below $10^{-3}$ requires $|\Delta_{i}|/\Omega\lesssim 0.01$ ($i=1,2$).
Over the simulated detuning range, DR gate shows an improvement on gate fidelity of almost or greater than one order of magnitude over the TO gate.

Note that we observe empirically that pulses suppressing only symmetric detuning error also frequently acquire substantial anti-symmetric-detuning robustness even when not explicitly enforced. We therefore only discuss robustness for arbitrary detuning error without a separate discussion on symmetric detuning-robust gate.

\section{Converting to erasure error}
\label{sec:4}

In fault-tolerant quantum error correction, both the physical error rate and the type of error determine the logical performance. 
Specifically, erasure errors that leave the computational subspace are directly detectable. After detecting the erasure errors their locations are known, therefore they are substantially easier to correct than Pauli errors, and a code operating under a channel dominated by erasures tolerates a higher physical error rate than the same code under depolarizing noise \cite{ grassl1997codes, barrett2010fault, Wu2022, Sahay2023, Kubica2023}. 

As a qubit candidate of growing interest for neutral atom quantum computation, the ytterbium-171 ($^{171}$Yb) atom is well suited to exploit this distinction by encoding the qubit in the long-lived, long-coherence $6s6p$ $^3$P$_0$ metastable states, from which the Rydberg transition is driven \cite{Ma2023, lis2023midcircuit, senoo2026high, Liu2026, zhang2026logical}. 
At the end of the gate the residual Rydberg population is converted out of the computational subspace, whose occupation can be detected by fluorescence or autoionization without disturbing metastable-state qubits \cite{Wu2022, Ma2023, scholl2023erasure}. A large fraction of gate errors is thereby heralded and converted into erasures rather than propagating as computational errors. 

\begin{figure}[t!]
    \centering
    \includegraphics[width=\linewidth]{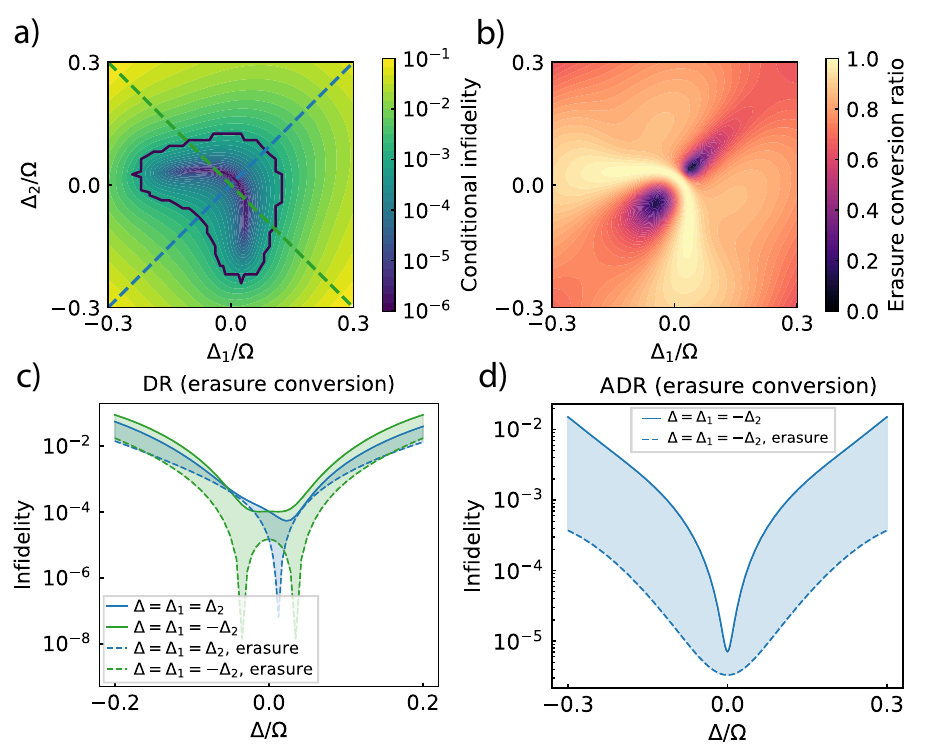}
    \caption{
    (a)(b) Conditional infidelity and erasure conversion ratio of the DR pulse of Fig. \ref{fig:2d-detuning}(c) under two-atom detuning, assuming complete conversion of residual Rydberg population into erasures. The contour encloses the region of simulated conditional fidelity above 99.9\%. The shapes of the high-conversion-ratio and high-conditional-fidelity regions are particular to this pulse solution and are not a generic feature of the robust-pulse construction. (c) Infidelity along the symmetric and antisymmetric diagonals of (a), before (solid) and after (dashed) erasure conversion. (d) Infidelity of the ADR pulse of Fig. \ref{fig:ADR}(b) under anti-symmetric detuning, before and after erasure conversion.
    }
    \label{fig:2d-erasure}
\end{figure}

We quantify this for the DR and ADR gates by assuming all residual population in $\ket{r}$ is converted to erasure errors, as is used in ref. \cite{Jandura2023},
and simulating the conditional gate fidelity of pulses in Fig. \ref{fig:ADR}(b) and Fig. \ref{fig:2d-detuning}(c). Figure \ref{fig:2d-erasure}(a-c) shows the conditional infidelity together with the associated conversion ratio, as a function of the two atom detuning. 
Figure \ref{fig:2d-erasure}(d) shows the effect of erasure conversion for ADR pulse under anti-symmetric detuning errors.
The conditional infidelity is far smaller than the raw infidelity over the relevant range of detunings, confirming that even after suppressing population leakage in optimization, a residual fraction of the infidelity originates from population remaining outside the computational subspace, i.e., in the Rydberg manifold, which can be converted into erasure errors. 

We find numerically that a single erasure-conversion step at the end of the sequence performs nearly as well as conversion after each $\sqrt{\text{CZ}}$. This follows from the structure of the echo: the interleaved X$\otimes$X pulses ensure that any given atom is excited to the Rydberg state in only one of the two $\sqrt{\text{CZ}}$ halves ($\ket{11}$ is mapped to the frozen $\ket{00}$, and $\ket{01}\leftrightarrow\ket{10}$ exchanges which atom is driven),
so the leaked amplitude of the first half is dynamically decoupled from the drive of the second and cannot coherently interfere with it. Any leaked population that does return to the computational subspace during the second $\sqrt{\text{CZ}}$ is separately suppressed, being of second order in the detuning by construction. The timing of the erasure conversion is therefore immaterial, and a single terminal round suffices.

\section{Conclusion}
\label{sec:5}

We have presented a family of Rydberg entangling pulses whose two-qubit entangling phase is robust to detuning errors on both atoms, together with an echoed sequence that converts them into a maximally-entangling ZZ gate. The central observation is that although no pulse can render the individual geometric phases insensitive to detuning \cite{Jandura2023}, the entangling angle $\theta_{11}-\theta_{10}-\theta_{01}$ can be made stationary. Targeting a $\sqrt{\text{CZ}}$ entangling angle and suppressing, in addition, the $\ket{r}$-leakage and $\ket{W_-}$-leakage amplitudes, confines all leading-order detuning errors to the single-qubit Z-rotations, which the interleaved X pulses then echo away. The resulting gate removes the leading-order response of the infidelity to arbitrary two-atom detuning, maintaining an infidelity below $10^{-3}$ over a range of single-atom detunings much wider than a comparable TO pulse.
In practice, such a fast interleaved echo sequence can be realized using stimulated Raman transitions, for instance, adding global Raman beams into the entangling gate zone for the zone architecture \cite{Bluvstein2022quantum, reichardt2024logical}.
We note that the optimization is performed in the frequency representation with an explicit bound on the modulation frequency under a certain Rabi frequency, so that every reported pulse respects the finite bandwidth of the modulator by construction (Appendix \ref{sec:FM}).

We have further shown that the residual error of these pulses is dominated by population remaining outside the computational subspace, which can be converted into heralded erasure error as in recent demonstrations in metastable $^{171}$Yb atoms \cite{Wu2022, Ma2023}.
Because the echo ensures that each atom is excited to the Rydberg state in only one of the two halves of the sequence, a single erasure-conversion step at the end suffices, with no need to interrupt the gate.

The anti-symmetric detuning case is relevant for ``drive-by'' gates~\cite{Lib2026,Xue2026DShor} in which a component of the laser k-vector is along the velocity of the moving atom. For ytterbium (rubidium) atoms moving at 1 m/s, the Doppler shift for the Rydberg transition is $\sim$3.3 MHz. This shift, corresponding to $2\Delta_a$ for the anti-symmetric detuning case, is too large to be effectively suppressed by the ADR gate described above when the Rabi frequency is small compared to the Doppler shift. However, the differential Doppler shift between a moving and a static atom could be compensated at 0th order by a differential light shift from the associated tweezer traps~\cite{Omran2019,Madjarov2020}, removing the average detuning. Although the fluctuation in this light shift due to tweezer intensity fluctuation and finite temperature would likely be sufficiently large to overwhelm the normal TO gates, our DR gate scheme is highly effective at suppressing such 1st order fluctuations around a zero-average detuning.  

Several extensions suggest themselves. First, the present pulses are optimized against detuning alone, whereas laser-amplitude inhomogeneity and pointing drift are comparably important in practice. Amplitude robustness enters the cost function through the same first-order perturbative structure used here, and pulses robust against both imperfections should be obtainable by combining the corresponding penalty terms, at the expense of a longer gate duration and hence increased Rydberg decay \cite{Jandura2023, Liu2026}. 
More broadly, one can apply a weaker set of constraints than a fully amplitude-robust gate that require only the two-qubit entangling phase and the leakage to be made insensitive to intensity fluctuations, with the residual single-qubit phases echoed away. We leave the design and characterization of such a gate to future work.
Second, the robustness conditions imposed here are first-order constraints, and the optimizer typically converges to a manifold of solutions satisfying them rather than to an isolated point. Characterizing this solution manifold through the Hessian of the cost function would identify the flat directions along which the pulse may be varied at no cost in robustness, and could be used to select, within the robust family, the pulse minimizing a secondary objective such as the Rydberg dwell time, the control bandwidth, or the higher-order detuning response \cite{Muniz2025, Liu2026}. Third, we have only explored the possibility of using two identical Rydberg pulse sequences to implement a single CZ gate. In general, the two Rydberg pulses could be different, as long as their sensitivities satisfy the appropriate robust condition, and their combined operation could also be other two qubit phase gates as well.

Finally, several effects excluded from the present model merit study. We have assumed infinite Rydberg blockade, and at finite blockade strength, the $\ket{11}$ sector acquires a coupling to $\ket{rr}$ whose leading effect is an interaction-induced shift of the $\ket{W_+}$ state. We note that a detuning-robust gate under fixed finite blockade can be numerically optimized (Appendix \ref{sec:finiteblockade}), but the contribution to the entangling phase from blockade strength fluctuation, unlike the detuning case, admits no cancellation between sectors. We have also assumed constant $|\Omega|$ during the gate time, while in real experiments Rydberg laser pulses have a finite switching on and off time, which can be easily implemented as a pre-designed amplitude modulation profile into the optimizer (Appendix \ref{sec:AM}). Likewise, we have assumed ideal, instantaneous single-qubit X-rotations and quasi-static detunings. A finite X-gate error reopens a fraction of the single-qubit phase error that the echo is designed to remove, and detuning fluctuations correlated on the timescale of the sequence would degrade the echo directly. Additionally, we check our gate fidelity under finite Rydberg state lifetime (Appendix \ref{sec:finitelifetime}), giving comparable performance to time-optimal gates especially after full erasure conversion, both with conditional fidelity $>99.9\%$. Establishing that these pulses preserve a high erasure bias under realistic noise is therefore a natural next step toward improved logical error rates in the metastable $^{171}$Yb architecture.

~

\textit{Note added}. --- During the completion of this manuscript, we became aware of a related work~\cite{EndresPC}.

~

\textit{Acknowledgments}. --- We acknowledge the Covey Lab for stimulating discussions. We acknowledge Bichen Zhang and Simon Evered for a critical reading of this manuscript.
J.P.C. acknowledges funding from the DOE Early Career Award from the Office of Nuclear Physics Quantum Horizons Program (award number DESC0025655); the Army Research Office (ARO awards W911NF-25-1-0156 and W911NF-25-1-0214); the NSF PHY Division (NSF award 2339487); and the U.S. Department of Energy, Office of Science, Q-NEXT National Quantum Information Science Research Center. 
The work is supported by the DOE Quantum Systems Accelerator (DE-SCL-0000121). 

~

\textit{Code availability.} --- Codes used for pulse optimization can be found in \cite{codes}.

\setcounter{section}{0}



\appendix
\renewcommand\appendixname{APPENDIX}
\renewcommand\thesection{\Alph{section}}
\renewcommand\thesubsection{\arabic{subsection}}

\setcounter{figure}{0}
\renewcommand{\thefigure}{S\arabic{figure}}

\section{Detuning robust gate under finite blockade}
\label{sec:finiteblockade}

The construction of Sec. \ref{Sec:2} assumes perfect blockade, in which $\ket{rr}$ is inaccessible and the $\ket{11}$ sector reduces to the two-level system $\{\ket{11}, \ket{W_+}\}$ driven at $\sqrt{2}\Omega$. In practice the blockade shift $B$ is finite, and $\ket{rr}$ acquires a small but nonzero amplitude during the gate. We show here that the design conditions generalize straightforwardly, and that a detuning-robust $\sqrt{\text{CZ}}$  can be obtained at fixed finite $B$ with no loss of robustness.

The states $\ket{00}$, $\ket{01}$ and $\ket{10}$ contain at most one Rydberg excitation and are unaffected, so their Hamiltonians and design conditions are unchanged. The $\ket{11}$ sector becomes a three-level ladder,

\begin{align}
H_{11}^{(0)} = \frac{\sqrt{2}\Omega}{2}[e^{i\phi}(| 11 \rangle \langle W_+ | + | W_+ \rangle \langle rr |) + \text{h.c.}] + B | rr \rangle  \langle rr |.
\end{align}

Requiring only the $\ket{11}$ trajectory to close at the gate time fixes only the first row and column of the unitary, so that $U_{11}^{(0)}(\tau)=e^{i\theta_{11}}\oplus V$ where $V$ is an arbitrary $2\times2$ unitary acting on $\left\{\ket{W_+},\ket{rr}\right\}$. We emphasize that $V$ does not need to be diagonal as the trajectory of $\ket{W_+}$ and $\ket{rr}$ do not necessarily need to close, the geometric phases of $\ket{W_+}$ and $\ket{rr}$ are therefore not individually well defined and the design conditions below are written so as not to require them.

Under a symmetric detuning error the perturbation acquires an additional diagonal term:
\begin{align}
    \Delta_sH_{11}^{(1)} = \Delta_s(\ket{W_+}\bra{W_+}+2\ket{rr}\bra{rr}).
\end{align}
With the extra coupling to $\ket{rr}$, the $\ket{r}$-leakage can come from residual population in either $\ket{W_+}$ or $\ket{rr}$. This leakage term now reads
\begin{align}
    \bra{f}U^{(1)}(\tau)\ket{11}
    =
    -i\int_0^\tau \left(\bra{f}VU_{11}^{(0)}(t)^{\dagger}H_{11}^{(1)}U_{11}^{(0)}(t)\ket{11}\right)\text{d}t
\end{align}
for $\ket{f}\in\left\{\ket{W_+},\ket{rr}\right\}$. 

The entangling-phase robustness condition acquires a corresponding term, changing eq. \eqref{robust} to

\begin{align}
\frac{\partial(\theta_{11}-2\theta_{10})}{\partial \Delta_s} =& 2\int_0^{\tau}|c_{10}(t)|^2\text{d}t - \int_0^{\tau}|c_{11}(t)|^2\text{d}t\nonumber\\
&-2\int_0^{\tau}|d_{11}(t)|^2\text{d}t=0,
\end{align}
with $d_{11}(t) = \bra{rr}U_{11}^{(0)}(t)\ket{11}$.

The antisymmetric-detuning condition, Eq. \eqref{W-}, is unchanged. The global drive does not couple $\ket{W_-}$ to $\ket{rr}$, and the differential detuning $\Delta_a$ acts only within the $\{\ket{W_+},\ket{W_-}\}$ manifold. The $\ket{W_-}$-leakage amplitude therefore still follows \eqref{W-} evaluated on the three-level trajectory.

\begin{figure}[t!]
    \centering
    \includegraphics[width=\linewidth]{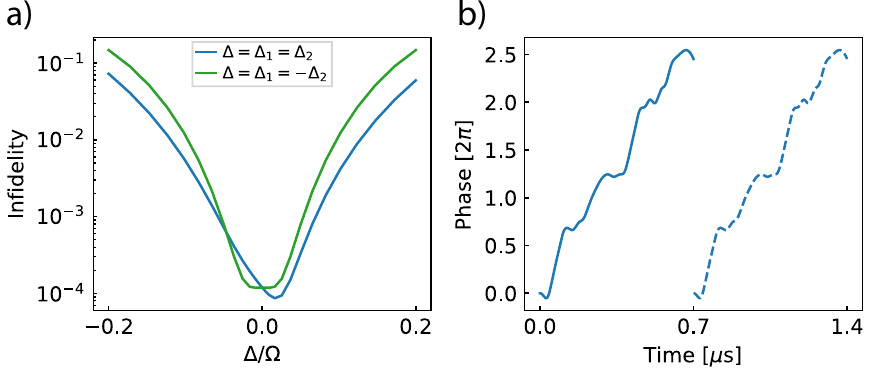}
    \caption{(a) Simulated gate infidelity under symmetric ($\Delta=\Delta_1=\Delta_2$) and antisymmetric ($\Delta=\Delta_1=-\Delta_2$) detuning for the finite-blockade DR pulses at $B/2\pi=60$ MHz. The leading-order response to detuning is removed for the DR pulse along both directions. (b) Optimized DR gate phase profile for $\Omega/2\pi=3$ MHz, $\tau=0.7$ $\mu$s and $B/2\pi=60$ MHz. The DR trace shows the two $\sqrt{\text{CZ}}$ halves of the echoed sequence (solid and dashed), each of the same duration.}
    \label{fig:finiteblockade}
\end{figure}

We optimized DR pulses under the modified conditions at $\Omega/2\pi=3$ MHz, $\tau=0.7$ $\mu$s, and $B/2\pi=60$ MHz. Figure \ref{fig:finiteblockade} shows the resulting infidelities under symmetric and antisymmetric detuning together with the optimized phase profiles, and the DR pulse retains the flat response established in Sec. \ref{sec:3}.

\section{Compatibility with finite amplitude ramp}
\label{sec:AM}

The pulses of Secs. \ref{sec:3}-\ref{sec:5} assume a constant Rabi frequency with instantaneous switching. In practice the amplitude is shaped by an acousto-optic modulator with a finite rise and fall time, and a smooth envelope is in any case desirable to avoid the broad spectral content and off-resonant transitions to neighboring Rydberg levels from the sharp switching. 

For a pre-designed amplitude profile, we can treat it as a fixed input for the optimizer by prescribing $\Omega(t)$ and optimizing only the phase $\phi(t)$ subject to the same cost function. 
Because the envelope is prescribed rather than optimized, it can equally be chosen to match the measured response of the modulator, or to match a certain amplitude modulation shape.
Precise characterization of the delivered amplitude and phase is achievable by heterodyne monitoring of the transmitted light \cite{Ma2023}.

\begin{figure}[t!]
    \centering
    \includegraphics[width=\linewidth]{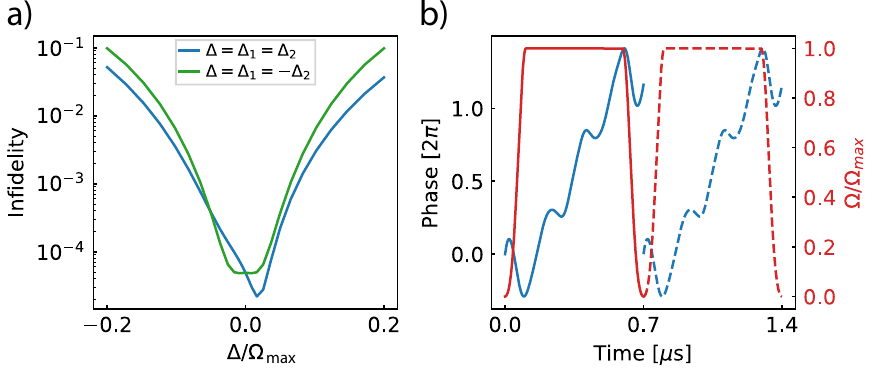}
    \caption{(a) Simulated gate infidelity under symmetric ($\Delta=\Delta_1=\Delta_2$) and antisymmetric ($\Delta=\Delta_1=-\Delta_2$) detuning for a DR pulse designed with a prescribed amplitude envelope. (b) Optimized phase profile (blue, left axis) and the prescribed Rabi frequency envelope (red, right axis), for $\Omega_{\text{max}}/2\pi=3.5$ MHz, $\sqrt{\text{CZ}}$  duration 0.7 $\mu$s, and rise/fall time 0.1 $\mu$s. Solid and dashed traces denote the first and second halves of the echoed sequence.}
    \label{fig:AM}
\end{figure}

We take a flat top at $\Omega_{\text{max}}/2\pi=3.5$ MHz with Blackman-shaped rise and fall of duration 0.1 $\mu$s \cite{blackman1958}. Figure \ref{fig:AM} shows the resulting pulse and its performance for a $\sqrt{\text{CZ}}$ of duration 0.7 $\mu$s. The simulation results show that the leading-order response to detuning is removed along both the symmetric and antisymmetric directions, with the infidelity remaining below $10^{-3}$ out to $\Delta/\Omega_{\text{max}}\approx0.06$, comparable to the constant Rabi frequency case.

\section{Frequency modulation bandwidth}
\label{sec:FM}

The modulator puts bandwidth limitations on the driving field modulation frequency $\omega(t)=\text{d}\phi/\text{d}t$. 
We therefore build our optimizer in the frequency representation: the control variables are the values of $\omega(t)$ on a coarse grid, from which $\phi(t)$ is recovered by integration and expanded onto the fine simulation grid through a smoothing spline. This makes the modulation frequency a direct optimization variable, so a bound on it is imposed as a simple box constraint. 
For the phase modulation profiles in the main text we optimize at a Rabi frequency of $\Omega/2\pi=3$ MHz, and cap the dynamical range of $\omega(t)/2\pi$ at 20 MHz throughout. 

\begin{figure}[t!]
    \centering
    \includegraphics[width=\linewidth]{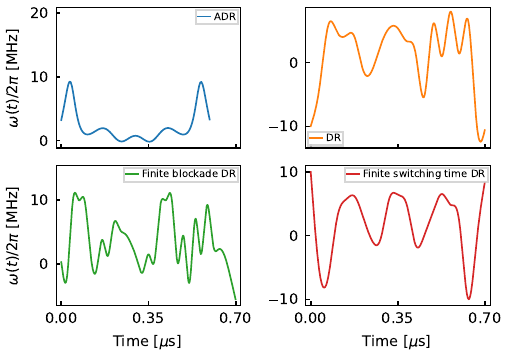}
    \caption{Modulation frequency $\omega(t)/2\pi$ of the four $\sqrt{\text{CZ}}$ pulses in Fig. \ref{fig:ADR}(b), \ref{fig:2d-detuning}(c), \ref{fig:finiteblockade}(b) and \ref{fig:AM}(b), plotted using the same frequency range. All lie within the 20 MHz bandwidth imposed during optimization.}
    \label{fig:FM}
\end{figure}

Figure \ref{fig:FM} shows the modulation frequency of the four pulses: the anti-symmetric-detuning-robust (ADR), detuning-robust (DR), finite-blockade DR, and finite-switching-time DR pulses. Instead of referring to the dimensionless unit $\Omega t$, here we specify the gate time $\tau$ to 0.6 $\mu$s for ADR gate and 0.7 $\mu$s for all other gates we compared here.
All frequency profiles lie within a range of 20 MHz. Its relevance is practical: the frequency response of an acousto-optic modulator rolls off over a bandwidth of order tens of MHz, and enforcing the bound during optimization rather than checking it afterward guarantees the pulses lie in the deliverable region by construction. 

We note that the ADR pulse was obtained with a separate optimizer that did not enforce this bound explicitly. Its modulation frequency nonetheless remains within the same range, so it too is compatible with a bandwidth-limited drive though this is not guaranteed in general. For the remaining pulses the frequency-domain parameterization with an explicit cap is the reliable route to a deliverable solution.

\begin{figure}[t!]
    \centering
    \includegraphics[width=\linewidth]{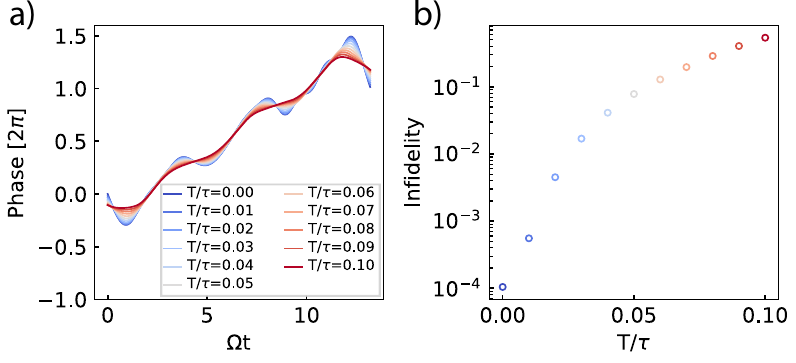}
    \caption{Effect of finite modulation bandwidth on the detuning-robust pulse, modeled by low-pass filtering the phase profile. (a) The phase $\phi(t)$ of the DR $\sqrt{\text{CZ}}$ pulse in fig. \ref{fig:2d-detuning}(b) after convolution with a moving-average window of width $T$, for windows from 0 to 10\% of the gate time $\tau$ (blue to red). Averaging suppresses the high-frequency content of $\phi(t)$ (equivalently, the high-frequency components of the modulation frequency $\omega(t)$) progressively rounding the sharpest features of the phase profile. (b) Corresponding error-free gate infidelity as a function of the window width $T/\tau$.}
    \label{fig:windowaverage}
\end{figure}

Another independent constraint concerns the rate of change of the modulation frequency, which is essentially limited by the beam size and the acoustic wave velocity in the acoustic-optical modulator crystal. 
As a simple model, we simulate the effect of finite modulation bandwidth by looking at the PM profile in fig. \ref{fig:2d-detuning}(b) after convolution with a moving-average window of various widths and checked the corresponding infidelity, as shown in fig. \ref{fig:windowaverage}. The infidelity remains below $10^{-3}$ for windows up to $\approx 1.5\%$ of the gate time and rises steeply beyond, quantifying the fidelity cost of removing high-frequency modulation content and thus the bandwidth the modulator must faithfully reproduce.

\begin{figure}[t!]
    \centering    \includegraphics[width=\linewidth]{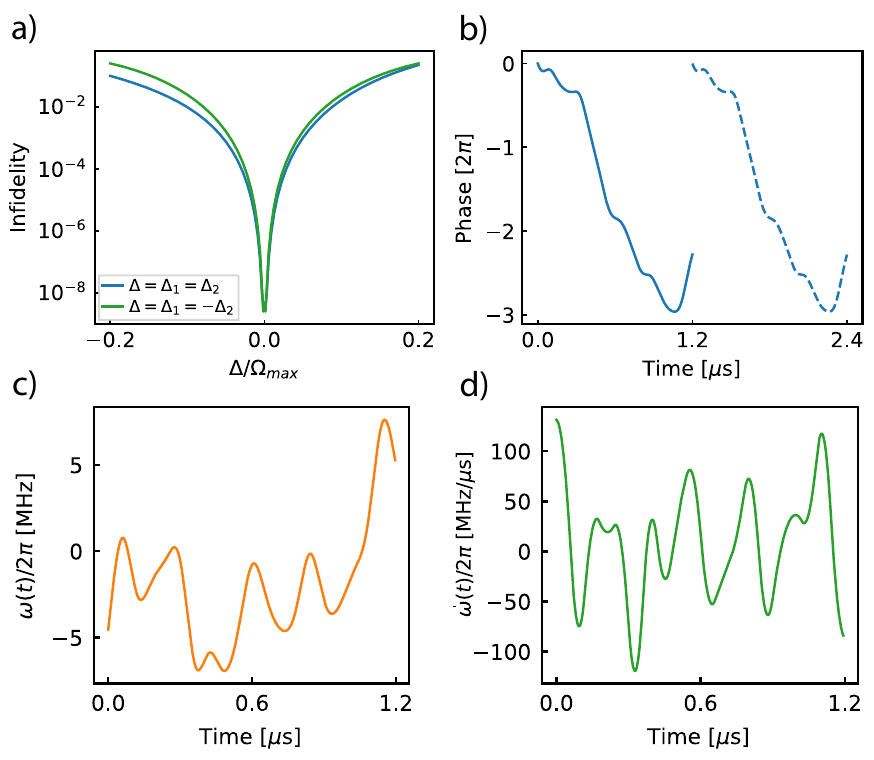}
    \caption{Detuning-robust $\sqrt{\text{CZ}}$ pulse designed under a bound on the chirp rate. (a) Simulated gate infidelity under symmetric ($\Delta=\Delta_1=\Delta_2$) and antisymmetric ($\Delta=\Delta_1=-\Delta_2$) detuning. The leading-order response is removed along both directions. (b) Optimized phase profile for $\Omega/2\pi=3$ MHz and $\tau=1.2$ $\mu$s with bandwidth constraints on $\omega$ and $\dot{\omega}$, with solid and dashed traces denoting the first and second $\sqrt{\text{CZ}}$ of the echoed sequence, each of duration 1.2 $\mu$s. (c) Modulation frequency $\omega(t)/2\pi$ and (d) its rate of change $\dot{\omega}(t)/2\pi$ for each $\sqrt{\text{CZ}}$ pulse, both constrained during optimization to $|\omega(t)|/2\pi\leq10$ MHz and $|\dot{\omega}(t)|/2\pi\lesssim 100$ MHz/$\mu$s.}
    \label{fig:DR_capped}
\end{figure}

We therefore impose a second constraint, $|\dot{\omega}|/2\pi\lesssim 100$ MHz/$\mu$s, on the optimization. Figure \ref{fig:DR_capped} shows the resulting pulse: panels (c) and (d) display the modulation frequency and its derivative, both within the imposed bounds, while panel (a) confirms that the robustness is preserved, with the leading-order response to detuning removed along both the symmetric and antisymmetric directions. The cost of the additional constraint is a longer gate: the 
$\sqrt{\text{CZ}}$ duration increases from 
0.7 to 1.2 $\mu$s under the same Rabi frequency.

\section{Finite Rydberg lifetime}
\label{sec:finitelifetime}

Unlike detuning errors, spontaneous decay from the Rydberg state cannot be suppressed by pulse shaping: the decay probability is set by the time the population spends in $\ket{r}$, which is bounded from below by the entangling phase the gate must accumulate. Robust pulses trade a longer Rydberg dwell for insensitivity to control errors, so it is important to establish that the robustness demonstrated in Secs. \ref{sec:3} and \ref{sec:4} is not obtained at a prohibitive cost in decay error. Here we quantify that cost, and show that it is largely recovered by erasure conversion.

We take the Rydberg-state lifetime $T_1=65$ $\mu$s and branching ratio into the (meta)stable levels measured for the $6s59s$ $^3$S$_1$ Rydberg state for $^{171}$Yb atoms \cite{Ma2023}, assume no Rydberg decay during single-qubit rotations and evaluate the decay error for the pulses of Fig. \ref{fig:2d-detuning}(c) under $\Omega/2\pi=3$ MHz, corresponding to 0.408 $\mu$s gate time for TO gate and 1.4 $\mu$s total Rydberg gate time for DR gate. We obtain a decay-limited fidelity of 99.46\% and 99.97\% after erasure conversion for DR gate, and decay-limited fidelity of 99.76\% and 99.99\% after erasure conversion for TO gate.

The echoed DR sequence occupies 3.5 times the duration of the TO gate, yet incurs only 2.25 times the decay error. The reason is that the two constructions distribute their Rydberg population differently. A $\sqrt{\text{CZ}}$  pulse and a TO pulse populate $\ket{W_+}$ comparably, but the $\sqrt{\text{CZ}}$  pulse spends only half as long in the singly-excited states $\ket{0r}$ and $\ket{r0}$, since it need accumulate only half the entangling phase. Under the echo, each atom is excited in only one of the two halves, so the $\ket{01}$ and $\ket{10}$ sectors accumulate the same total Rydberg dwell as a single TO gate, while the $\ket{11}$ sector accumulates roughly twice. The net cost is therefore significantly less than a straightforward scaling on gate time.

Because the qubit is encoded in the metastable $^3$P$_0$ manifold, decay from $\ket{r}$ leaves the computational subspace and is detectable \cite{Wu2022, Ma2023}, so these errors are precisely those that erasure conversion is designed to capture. Assuming complete conversion of the residual Rydberg population, both gates exceed 99.9\% conditional fidelity, and the difference between them becomes negligible. The robustness of the DR gate is thus obtained at essentially no cost in conditional fidelity, which is the quantity that governs the logical error rate in an erasure-biased architecture \cite{Jandura2023}.

\section{Robustness against time-dependent detuning error}
\label{sec:timedependentdetuning}

The sequence is designed to suppress a constant detuning error, such as that from a DC electric field, but not a time-varying one. We Taylor expand the detuning about the sequence time and keep the leading term; the constant part is removed by the detuning-robust construction, leaving a linear drift $\Delta=\Delta' t$. To first order, the $\ket{11}$-sector phase acquired by a single Rydberg pulse beginning at time $t_{\text{start}}$ is 
\begin{align}
    \delta\theta_{11}^{(1)}(t_\text{start}) = \frac{\partial\theta_{11}}{\partial\Delta}\,\Delta'\,t_\text{start} + \frac{\partial\theta_{11}}{\partial\Delta'}\,\Delta',
\end{align}
where $\partial\theta_{11}/\partial\Delta$ and $\partial\theta_{11}/\partial\Delta'$ are the sensitivities to a constant detuning and to a linear ramp. Because the echo swaps $\ket{11}\leftrightarrow\ket{00}$ between the two halves, the residual entangling-phase error is the difference of this quantity evaluated at the two pulse start times, separated by $t_{\text{sep}}$. The linear-ramp sensitivity $\partial\theta_{11}/\partial\Delta'$ is common to both halves and cancels; what remains is
\begin{align}
    \delta\theta^{(1)}_{00\text{-}11} = \frac{\partial\theta_{11}}{\partial\Delta}\Delta't_\text{sep}.
\end{align}
The echo thus converts a time-varying detuning into an effective constant detuning mismatch $\Delta't_{\text{sep}}$ between the two halves, to which the gate responds through its constant-detuning sensitivity. Since $\partial\theta_{11}/\partial\Delta\neq 0$, this residual cannot vanish within the simple echo. It could be suppressed by a more elaborate sequence, for instance, two different Rydberg pulses whose detuning-ramp sensitivities are different, or additional echo steps. We leave it to future work.


\bibliographystyle{apsrev4-2}
\bibliography{refs}

\end{document}